\documentclass[9pt,conference]{IEEEtran}
\IEEEoverridecommandlockouts
\usepackage{cite}
\usepackage{amsmath,amssymb,amsfonts}

\usepackage[sort&compress, numbers]{natbib} 
\usepackage{algorithm}
\usepackage{algpseudocode}
\usepackage{graphicx}
\usepackage{textcomp}
\usepackage{xcolor}
\usepackage{subcaption}
\usepackage{tikz}
\usepackage[T1]{fontenc}
\usepackage{times}
\usepackage{hyperref}
\usepackage[labelformat=simple]{caption}
\def\BibTeX{{\rm B\kern-.05em{\sc i\kern-.025em b}\kern-.08em
    T\kern-.1667em\lower.7ex\hbox{E}\kern-.125emX}}

\newcommand{\method}{HD-MoE}

\definecolor{color1}{RGB}{78, 121, 167}   
\definecolor{color2}{RGB}{242, 142, 43}   
\definecolor{color3}{RGB}{225, 87, 89}    
\definecolor{color4}{RGB}{89, 161, 79}    
    
\begin{document}

\title{\method: Hybrid and Dynamic Parallelism for Mixture-of-Expert LLMs with 3D Near-Memory Processing}

\author{\IEEEauthorblockN{Haochen Huang$^{1,2\dagger}$, Shuzhang Zhong$^{1,2,3,4\dagger}$, Zhe Zhang$^{3,4}$, Shuangchen Li$^{3,4}$}
\IEEEauthorblockN{Dimin Niu$^{3,4}$, Hongzhong Zheng$^{3,4}$, Runsheng Wang$^{2,5,6}$, Meng Li$^{1,2,6*}$}
\IEEEauthorblockA{$^1$Institute for Artificial Intelligence, Peking University, Beijing, China}
\IEEEauthorblockA{$^2$School of Integrated Circuits, Peking University, Beijing, China}
\IEEEauthorblockA{$^3$DAMO Academy, Alibaba Group, Beijing, China}
\IEEEauthorblockA{$^4$Hupan Lab, Hangzhou, China}
\IEEEauthorblockA{$^5$Institute of Electronic Design Automation, Peking University, Wuxi, China}
\IEEEauthorblockA{$^6$Beijing Advanced Innovation Center for Integrated Circuits, Beijing, China}

\thanks{
This work was supported in part by NSFC under Grant 62495102 and Grant 92464104, in part by the National Key Research and Development Program under Grant 2024YFB4505004, in part by Beijing Municipal Science and Technology Program under Grant Z241100004224015, and in part by 111 Project under Grant B18001.

$^\dagger$Equal Contribution
$^*$Corresponding author: meng.li@pku.edu.cn}
}


\maketitle

\begin{abstract}
Large Language Models (LLMs) with Mixture-of-Expert (MoE) architectures achieve superior model performance with reduced computation costs, but at the cost of high memory capacity and bandwidth requirements. Near-Memory Processing (NMP) accelerators that stack memory directly on the compute through hybrid bonding have demonstrated high bandwidth with high energy efficiency, becoming a promising architecture for MoE models. However, as NMP accelerators comprise distributed memory and computation, how to map the MoE computation directly determines the LLM inference efficiency. Existing parallel mapping strategies, including Tensor Parallelism (TP) and Expert Parallelism (EP), suffer from either high communication costs or unbalanced computation utilization, leading to inferior efficiency. The dynamic routing mechanism of MoE LLMs further aggravates the efficiency challenges. Therefore, in this paper,  we propose \method~to automatically optimize the MoE parallel computation across an NMP accelerator. \method~features an offline automatic hybrid parallel mapping algorithm and an online dynamic scheduling strategy to reduce the communication costs while maximizing the computation utilization. With extensive experimental results, we demonstrate that \method~ achieves a speedup ranging from 1.1$\times$ to 1.8$\times$ over TP, 1.1$\times$ to 1.5$\times$ over EP, and 1.0× to 1.4× over the baseline Hybrid TP-EP with Compute-Balanced parallelism strategies.


\end{abstract}

\section{Introduction}

Mixture-of-Experts (MoE) has become a widely adopted architecture for Large Language Models (LLMs) \cite{deepseekv3,jiang2024mixtral}. By selectively activating only a small subset of experts, MoE significantly reduces the computational demands while maintaining model capacity. However, the sparse activation mechanism exacerbates the memory-bound problem, particularly on edge devices with limited memory bandwidth and small batch sizes.

The recent emerging 3D Near-Memory Processing (NMP) architectures seem to be a promising solution for memory-bound problems \cite{memorywall,Near-memory-FPGAs}. 3D NMP vertically stacks DRAM dies directly on top of logic dies using high-bandwidth interconnects. In contrast to conventional von Neumann architectures, the vertical stacking of 3D NMP allows multiple memory banks to be accessed independently and in parallel, enabling fine-grained, high-throughput data access. This makes 3D NMP particularly suitable for MoE inference workloads.

\begin{figure}[!tb]
    \centering
    \includegraphics[width=\linewidth]{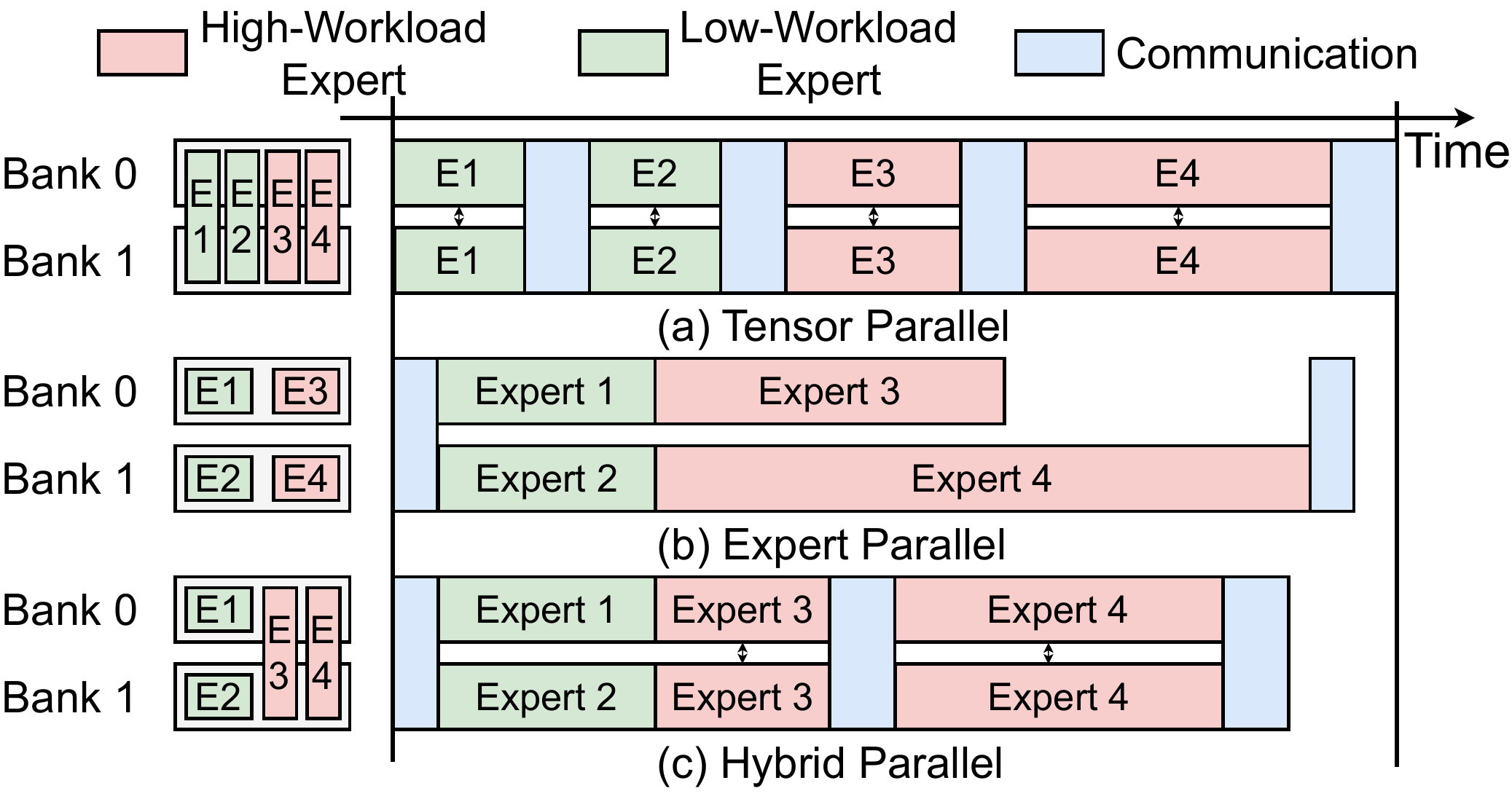}
    \caption{Expert Deployment and Computation Timeline in (a) Tensor Parallel, (b) Expert Parallel, and (c) Hybrid parallel.}
    \label{fig:intro}
\end{figure}

While MoE's bandwidth efficiency makes it suitable for 3D NMP deployment, the architectural shift from GPU-style shared memory to distributed NoC-based designs introduces new mapping challenges. The distributed nature of 3D NMP, with its bank-local memory organization, requires careful co-design of expert parallelism and communication routing strategies to maintain performance. As illustrated in Figure~\ref{fig:intro}, current approaches employ either Tensor Parallelism \cite{TP} (TP) or Expert Parallelism \cite{lepikhin2020gshard} (EP): TP distributes each expert's parameter tensor across banks while EP assigns complete experts to different banks. This presents a fundamental trade-off: TP achieves better workload balance but incurs substantial all-reduce communication overhead, whereas EP minimizes communication but suffers from workload imbalance due to varying expert utilization. 

Previous works on GPU clusters have explored combining EP with replication of frequently activated experts to achieve both workload balance and low communication overhead. This method has been adopted by DeepSeek-AI to deploy its DeepSeek-R1 model \cite{deepseekv3}. However, this approach is impractical for 3D NMP due to its limited memory capacity. Furthermore, the dynamic and imbalanced nature of expert activation patterns significantly complicates mapping and scheduling decisions, requiring more sophisticated optimization strategies tailored to the constraints of 3D NMP architectures.

To address the challenge of dynamic expert activation, several studies focusing on offloading scenarios have investigated dynamic scheduling of experts \cite{AdapMoE, lin2024task, zhang2024daop, tang2024hobbit}. In these scenarios, experts are stored in secondary storage, with on-demand loading becoming the primary bottleneck. These studies demonstrate that MoE models often exhibit high activation similarity between adjacent layers, which can be exploited for prefetching to alleviate the on-demand loading overhead.

In light of these challenges and opportunities, we propose \textbf{HD-MoE}, a hybrid and dynamic parallelism framework designed for MoE inference on 3D NMP architectures. Reducing latency on distributed systems requires both \textbf{balancing computation utilization} and \textbf{minimizing communication cost}, while also addressing memory limitations. To achieve this, HD-MoE adopts a hybrid parallelism approach, as illustrated in Figure~\ref{fig:intro}(c). Experts with low activation frequency are mapped using Expert Parallelism to minimize communication overhead, while high-frequency experts utilize Tensor Parallelism to maximize computational resource utilization. Additionally, HD-MoE incorporates an online dynamic expert placement strategy to mitigate the impact of the dynamic activation pattern. The related codes can be accessed at \url{https://github.com/angerybob/HD-MoE}.

The key contributions of HD-MoE are summarized as follows:
\begin{itemize}
    \item \textbf{Performance Analytical Model.} We develop a unified performance analysis framework applicable to diverse hardware configurations and parallelism strategies.
    \item \textbf{Automated Hybrid Parallelism.} We propose an efficient placement strategy searching method that combines TP and EP to optimize computation and communication overheads.
    \item \textbf{Dynamic Placement.} We introduce a dynamic expert placement strategy, which adjusts expert deployment in real-time based on the inference workload, ensuring optimal performance even in different inference scenarios.
    \item We conduct extensive experiments to validate our approach, demonstrating significant improvements in both TBT latency and speedup compared to baseline methods.
\end{itemize}

\section{Background}

\subsection{Mixture-of-Experts Models}

\begin{figure}
    \centering
    \includegraphics[width=\linewidth]{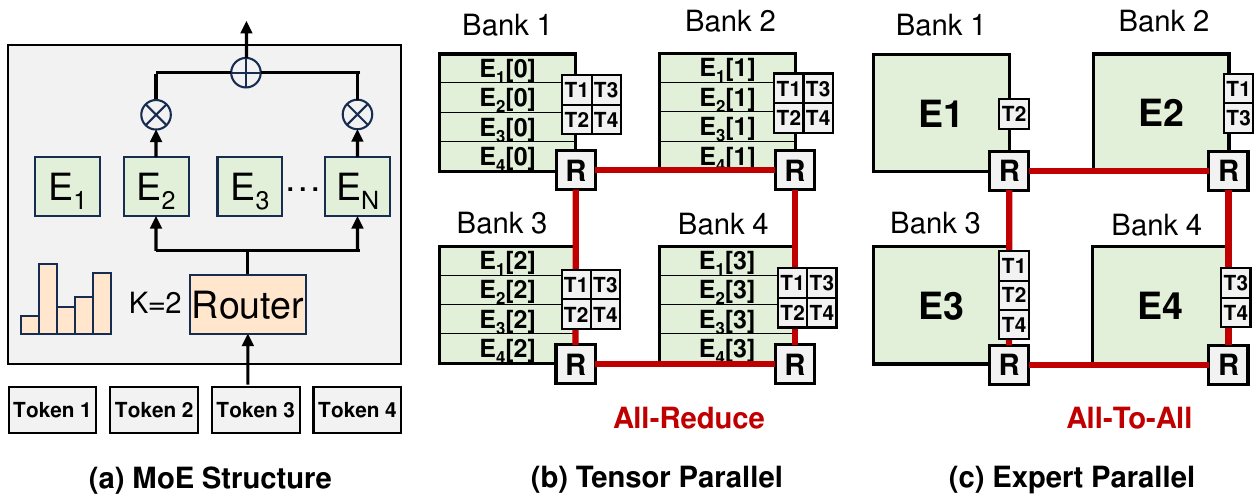}
    \caption{MoE structure and two parallel strategies}
    \label{fig:background}
\end{figure}

Mixture-of-Experts (MoE) improves scalability by activating a small subset of specialized subnetworks per input, reducing computation and memory costs. A gating mechanism selects experts per token, enabling efficient large-scale and multi-task learning. Recent MoE models such as Mixtral~\cite{jiang2024mixtral}, Qwen~\cite{qwen}, and DeepSeek~\cite{deepseekv3, deepseekv2} have shown significant performance and scalability gains.

MoE optimization has been explored across inference and training. Offloading methods target hybrid CPU-GPU deployment for throughput efficiency~\cite{AdapMoE,HybriMoE}; serving-oriented works improve expert scheduling and placement~\cite{eps,moetuner,SCmoe}; and training strategies integrate data, tensor, and expert parallelism to enhance scalability~\cite{lepikhin2020gshard,Switch,FasterMoE,netmoe}. These approaches highlight the need for adaptable solutions under diverse constraints in memory, computation, and communication.

\subsection{Near-Memory Processing Architectures}

Traditional von Neumann architectures, including GPUs, face increasing limitations from the memory wall, particularly in sparse, bandwidth-intensive models like MoEs. Processing-in-Memory (PIM) alleviates memory latency by colocating computation with data~\cite{kim2023samsung}, but its low compute density, due to logic-storage contention, limits scalability. In contrast, 3D Near-Memory Processing (NMP) vertically integrates DRAM and logic dies to provide high bandwidth and moderate compute density, making it more suitable for bandwidth-bound workloads.

An example is \textbf{Hybrid Bonding DRAM}~\cite{HB1,HB2,similarity,li2025h2}, which employs fine-pitch vertical interconnects for high-bandwidth links between memory and logic, supporting large-scale sparse models efficiently.

\subsection{Distributed Inference Strategies}

Distributed inference on 3D NMP architectures requires efficient parallelism to utilize compute and memory resources, with Tensor Parallelism (TP) and Expert Parallelism (EP) as two primary strategies. As shown in Fig.~\ref{fig:background}b, TP splits expert parameters across processing elements (PEs), achieving balanced computation but incurring costly all-reduce communication that scales with batch size. EP (Fig.~\ref{fig:background}c) assigns entire experts to separate PEs, reducing communication per token but causing load imbalance and irregular all-to-all traffic due to sparse, dynamic expert activation.

Fig. \ref{fig:background} compares the two strategies. Our approach combines them in a hybrid framework to optimize MoE inference on 3D NMP systems, efficiently balancing compute and communication overhead.

\section{Motivation}

\begin{table}
    \centering
    \caption{Comparison of MoE Deployment Scenarios on Different Hardware Architectures.}
    \begin{tabular}{c|c|c|c|c} \hline \hline
          &Scenario
&  Memory&  Capacity& Parallelism
\\ \hline 
         GPU &Edge Inference&  Shared&  Low& TP
\\ \hline 
         Cluster &Cloud Serving&  Distributed&  High& EP+Replicate
\\ \hline 
         3D NMP &Edge Inference&  Distributed&  Low& Hybrid
\\ \hline \hline
    \end{tabular}
    \label{tab:comparison}
\end{table}

\textbf{Challenge 1: MoE Deployment Challenges Unique to 3D NMP Architectures.} 3D NMP architectures offer high bandwidth but lack shared memory across compute units, resulting in a distributed execution environment distinct from traditional GPUs or clusters.

In single-GPU systems, global memory and shared L2 cache enable low-latency communication among SMs, making Tensor Parallelism (TP) effective for balancing workloads. In contrast, cluster deployments favor Expert Parallelism (EP) to reduce inter-node communication, often replicating hot experts to improve load balance with minimal memory overhead—especially feasible when KV cache dominates memory use. 

However, these strategies are ill-suited for 3D NMP. As shown in Table~\ref{tab:comparison}, TP incurs prohibitive communication overhead due to limited NoC bandwidth, while EP with replication exceeds the memory budget. This calls for a hybrid parallelism strategy tailored to 3D NMP constraints, balancing communication cost, workload skew, and tight storage capacity for scalable MoE inference.

\begin{figure}[!tb]
\centerline{\includegraphics[width=\linewidth]{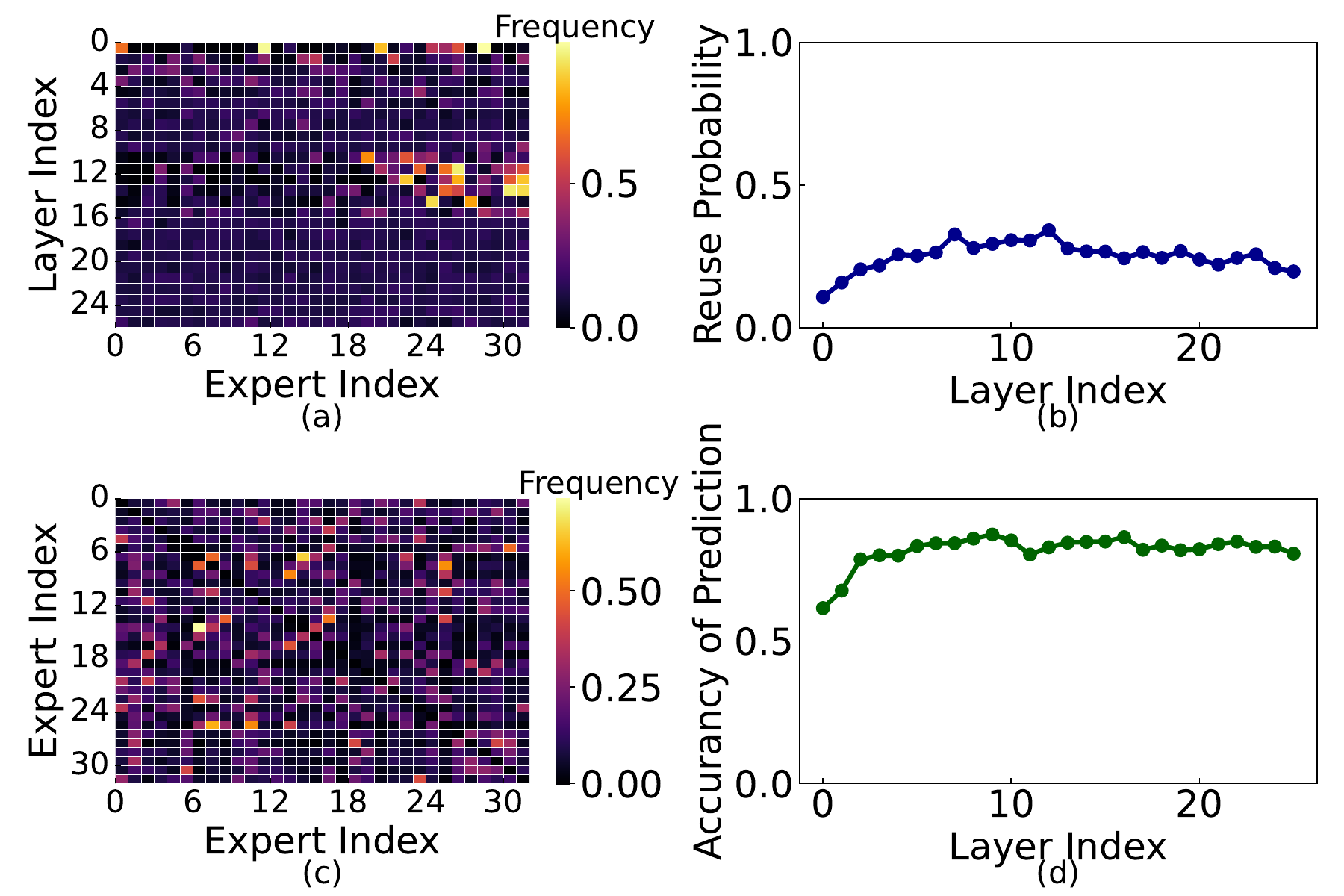}}
\caption{(a) Expert Activation Frequency, (b) Expert Activation Overlap Between Adjacent Iterations, (c) Expert Routing Affinity, The value at position (i,j) in the figure represents the conditional probability of expert j being activated, given that expert i is activated, (d) Prediction Accuracy of Expert Activation.}
\label{fig:motivation}
\end{figure}

\textbf{Challenge 2: Imbalanced and Dynamic Expert Activation of MoE.}
The expert selection frequency in MoE is inherently uneven, leading to significant workload imbalance. For instance, as shown in Fig.~\ref{fig:motivation}(a), nearly half of the tokens in certain layers of Qwen2 converge to a single expert, while others remain underutilized \cite{yang2024qwen2technicalreport}. This imbalance creates both computational and communication inefficiencies. Overloaded experts become performance bottlenecks, while underutilized experts waste computing resources. Furthermore, the skewed workload triggers inefficient all-to-all communication patterns in 2D mesh networks, increasing latency and reducing effective bandwidth. Together, these effects significantly degrade system performance.

Dynamic expert activation exacerbates these challenges. As illustrated in Figure~\ref{fig:motivation}(b), expert selection exhibits high variability across iterations, with some experts experiencing heavy utilization in one iteration and minimal usage in the next. Thus, an efficient MoE deployment must address both static imbalance and dynamic variability to optimize compute and communication efficiency.

\textbf{Opportunity 1: Activation Dependency in Expert Groups.} Despite MoE activation imbalance, experts show strong co-activation affinity. For example, when expert A is activated, expert B is more likely to be activated, forming a distinct co-activation pattern (Fig.~\ref{fig:motivation}(c)). This pattern enables optimization by co-locating frequently co-activated experts in 3D NMP, reducing data movement, optimizing communication, and balancing device loads. The key challenge is leveraging activation dependencies and 3D NMP architecture features to address load imbalance and communication latency, improving MoE inference performance.

\textbf{Opportunity 2: Expert Activation Prediction.} MoE models exhibit strong activation similarity between adjacent layers due to residual connections, enabling accurate prediction of future expert activations. As shown in Figure~\ref{fig:motivation}(d), the gate functions of subsequent layers achieve high prediction accuracy, providing a promising approach to mitigate the challenges of dynamic activation patterns. By leveraging these predictions, we can proactively optimize resource allocation, reduce communication overhead, and improve load balancing in distributed systems.

\section{HD-MoE Design}

\subsection{Overview}

\begin{figure}[!tb]
\centerline{\includegraphics[width=\linewidth]{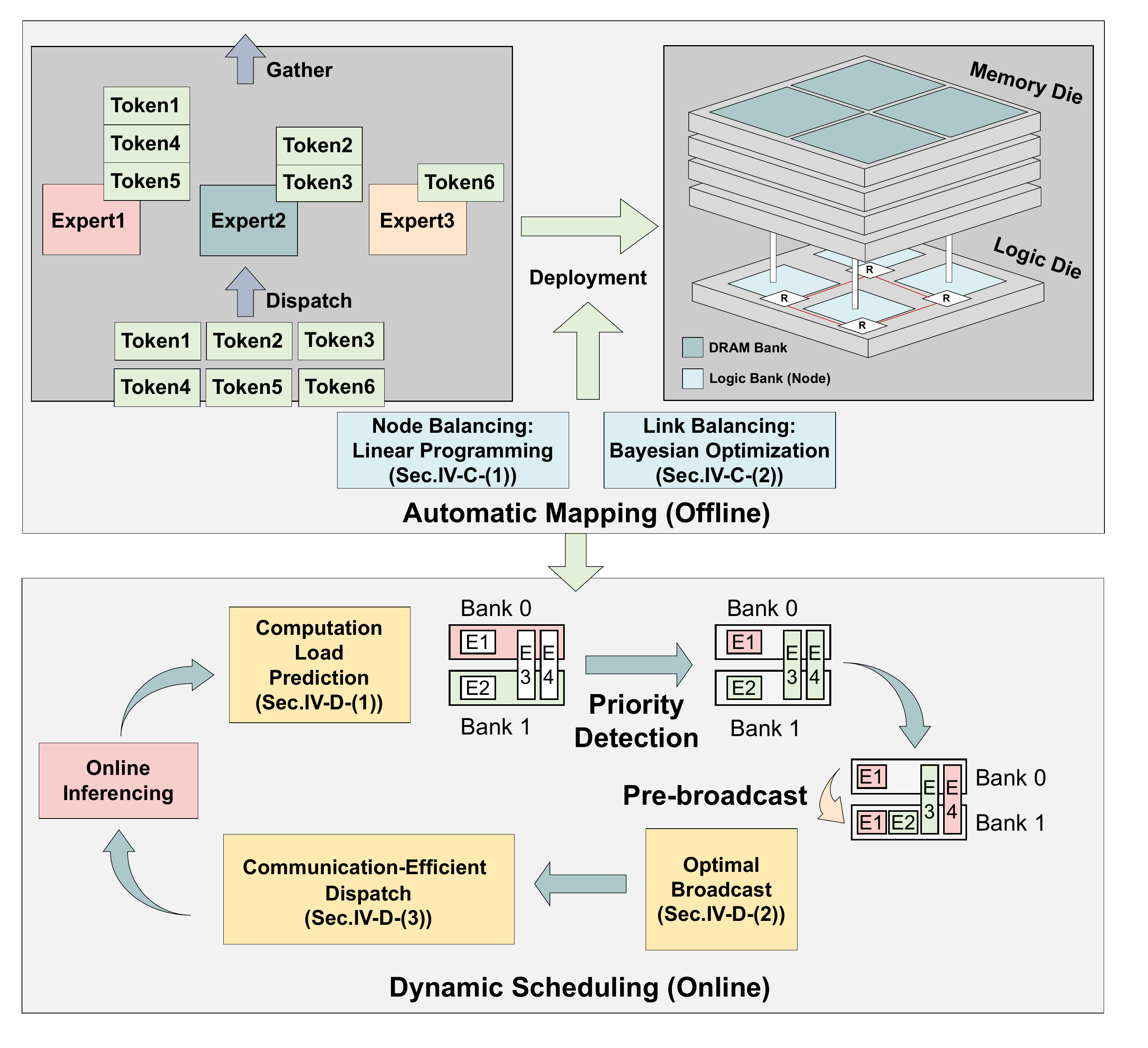}}
\caption{Overview of HD-MoE}
\label{overview}
 \end{figure}

Fig. \ref{overview} provides an overview of the HD-MoE framework, which consists of an offline mapping phase and an online inference phase. The offline phase involves automated hybrid expert mapping on 3D NMP with Node Balance (Sec. \ref{node balance}) and Link Balance (Sec. \ref{link balance}) optimization techniques. During online inference, dynamic scheduling is employed to predict computation load, prioritize experts (Sec. \ref{prediction}), and pre-broadcast the expert with highest load (Sec. \ref{broadcast}), ensuring efficient resource utilization without inducing additional communication overhead (Sec. \ref{communication-eff}).

\subsection{Performance Analytical Model}
We first build a performance analytical model to estimate the total inference cost, including computation and communication estimation.

\subsubsection{Computation Overhead Modeling} 
The computation time $t_{\text{comp}}$ is determined by the maximum load across all computing nodes, reflecting expert utilization imbalance. For each node $c$, the load depends on the placement matrix $P_{ic}$, where $P_{ic}$ denotes the proportion that expert $i$ is assigned to node $c$. The explanation of other parameters is listed in the table~\ref{parameters}. The formula is defined as:
\begin{align}
t_{\text{comp}} = \max_{c} \left\{ \frac{\sum_{i=0}^{E-1} P_{ic} f_i B \cdot 2h \cdot IS}{\text{comp}} \right\}\label{comp_load}
\end{align}
Here, $2h \cdot IS$ is the computation volume per token, $f_iB$ denotes the number of tokens that activate expert $i$, and $P_{ic}f_iB$ is the effective number of tokens that node $c$ needs to process.

We emphasize that the variables $P_{ic}$ are continuous rather than binary, allowing our hybrid placement strategy to partially distribute expert computation across multiple nodes (similar to Tensor Parallelism). This design improves deployment flexibility, alleviates hotspots, and enhances compute balance.

However, this introduces a trade-off with communication overhead. Splitting experts across nodes requires data transfers, which can increase both transfer volume and communication irregularity. Balancing computation load and communication overhead is crucial for optimal MoE performance on 3D PNM architectures.

\subsubsection{Communication Overhead Modeling}\label{comm_model}

We adopt a discrete-event simulation framework to model irregular all-to-all communications in 2D mesh architectures, extracting the total schedule time as communication overhead.

\textbf{Discrete-Event Simulation for Accurate Latency Estimation.} We build a discrete-event simulator to model the latency of irregular all-to-all communication in 2D mesh networks, which mainly consists of the following parts: 
\begin{itemize}
    \item \textbf{Communication Task Generation:} For each token’s activated expert group, we first map experts to their physical nodes (src) and identify a target node (dst) for aggregation (e.g., randomly choose a node that is activated by this token).

    \item \textbf{XY Routing Path Calculation:} We use a cached XY routing algorithm to generate minimal-hop paths (Manhattan distance) between source and destination nodes. 

    \item \textbf{Event Scheduling and Link Management:} Communication tasks are split into chunks and scheduled in a priority queue. Transmission time is calculated based on available bandwidth and link occupancy. A link schedule dictionary tracks link availability, ensuring tasks are scheduled promptly and avoiding collisions.

\end{itemize}

\textbf{Linear Approximation for Optimization.} To enable efficient deployment optimization via linear programming (LP), we approximate the communication latency using a node-traffic model: 
\begin{align}
\hat{t}_{\text{comm}} = \frac{4}{\text{BW}} \max_{c} \left\{ \sum_{g \in G} \left( \prod_{i \in g} \lceil P_{ic} \rceil \right) f_g B h \right\}
\end{align}
Here, $\lceil P_{ic} \rceil$ indicates whether expert $i$ is activated on node $c$, and $\prod_{i \in g} \lceil P_{ic} \rceil$ checks if any expert in group $g$ is placed on node $c$. If any expert in group $g$ is placed, the data volume for transfer is $4 f_g B h$, assuming FP32 representation. The total communication volume that node $c$ needs to send is given by $\sum_{g \in G} \left( \prod_{i \in g} \lceil P_{ic} \rceil \right)4 f_g B h$.

\begin{table}[!tb]
\centering
\caption{Explanation of Parameter Meanings.}
\begin{center}
\begin{tabular}{c|c}

\hline\hline
\textbf{Parameter} & \textbf{Meaning} \\
\hline
$t_{\text{comp}}$ & Computation time \\
\hline
$t_{\text{comm}}$ & Communication time \\
\hline
$\hat{t}_{\text{comm}}$ & Linear Approximation for communication time\\
\hline
$c$ & Index of computing nodes \\
\hline
$E$ & Total number of experts \\
\hline
$e$ & Number of experts activated per token\\
\hline
$P_{ic}$ & Proportion of expert $i$ placed on node $c$ \\
\hline
$f_i$ & Activation frequency of expert $i$ \\
\hline
$B$ & Batch size \\
\hline
$h$ & Hidden dimension \\
\hline
$IS$ & MoE intermediate size \\
\hline
$D$ & Number of nodes \\
\hline
$\text{comp}$ & Computational throughput of a node \\
\hline
$\text{BW}$ & NoC link bandwidth \\
\hline
$G$ & Expert groups \\
\hline
$g$ & Index of expert groups \\
\hline
$f_g$ & Co-activation frequency of expert group $g$ \\
\hline\hline
\end{tabular}
\label{parameters}
\end{center}
\end{table}

\begin{figure}[!tb]
\centerline{\includegraphics[width=0.9\linewidth]{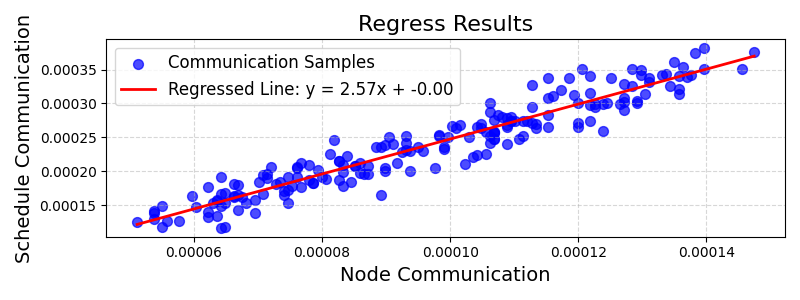}}
\caption{The fitting results illustrating the relationship between schedule-based communication mechanisms and node communication patterns ($R^2=0.96$).}
\label{communicaton}
\end{figure}

We empirically validate the accuracy of the proposed linear approximation model. By comparing the estimated communication latency $\hat{t}_{\text{comm}}$ with the latency values obtained from simulation $t_{\text{comm}}$, we observe a strong linear correlation between the two. The fitting results are shown in Fig.~\ref{communicaton}. 

As a result, the relationship can be approximated as:
\begin{align}
 t_{\text{comm}} = \gamma \hat{t}_{\text{comm}}
\end{align}
where $\gamma$ is a scaling coefficient determined through linear regression. Experimental results show that the coefficient of determination ($R^2$) consistently exceeds 0.9 across various scenarios, confirming the model’s reliability in estimating communication overhead for LP.

In structured communication patterns like ring-based all-reduce \cite{ring}, commonly used in TP training and inference, the communication algorithm is regular and deterministic. In such cases, total communication time nearly equals per-node communication time due to the algorithm's balanced nature.
\begin{align}
 t_{\text{comm}} \approx  \hat{t}_{\text{comm}} \approx \frac{4Bh}{\text{BW}}
\end{align}
This implies that $\gamma = 1$ for ring all-reduce, demonstrating that our linear communication model is not only accurate for irregular all-to-all traffic, but also directly applicable to structured communication paradigms such as tensor parallel all-reduce.

We validate our analytical model by comparing its results with the widely used distributed deep learning simulator ASTRA-sim \cite{9238637} for the ring all-reduce operation. As shown in the table \ref{tab:astra_sim}, the latency predictions from the analytical model closely match the simulation results, demonstrating strong alignment between the two. This confirms that our analytical model provides a reliable estimate of latency and can be effectively used for performance prediction and optimization in similar distributed systems.

\begin{table}[!tb]
\centering
\caption{Comparison between the analytical modeling and ASTRA-sim-based simulation results for Ring-AllReduce.}

\begin{center}
\resizebox{0.5\textwidth}{!}{ 

    \begin{tabular}{c|c|c|c}
    \hline\hline
    \textbf{Latency} & \textbf{Bandwidth} & \textbf{Analytical Results} & \textbf{Simulation Results} \\
    \hline
    0.1 us & 25 Gb/s & 673 us & 668 us \\
    \hline
    5 us & 25 Gb/s & 751 us & 879 us \\
    \hline
    0 & 20 Gb/s & 671 us & 692 us \\
    \hline\hline
    \end{tabular}}
    \label{tab:astra_sim}
\end{center}
\end{table}

\subsection{Optimal Placement Strategy Searching}

We propose a two-stage \textbf{Node-Link Balance Co-optimization} strategy for efficiently deploying MoE models on 3D NMP architectures. The placement problem is divided into a logical optimization stage that balances computational load and reduces communication volume, and a physical mapping stage to minimize link-level congestion. This separation of logical workload and physical interconnects simplifies the placement problem, as detailed in the following stages.

\subsubsection{Node Balance Optimization via Linear Programming}
\label{node balance}

In the first stage of our co-optimization framework, we focus on balancing computational and communication workloads across logical compute clusters, abstracting away their physical topology. Given the large number of variables and the combinatorial nature of the expert placement problem, manual tuning becomes infeasible. Therefore, we adopt a linear programming (LP) formulation to enable automated and scalable optimization across diverse hardware configurations and inference scenarios. The LP model simultaneously considers computation bottlenecks and approximated communication costs using the linear estimator $\hat{t}_{\text{comm}}$ derived earlier. The LP Optimization Problem is modeled as follows:

The notations of the LP formulation are defined in Table~\ref{parameters}. Among them, the continuous variables $P_{ic} \in [0, 1]$ represent the proportion of expert $i$’s workload assigned to cluster $c$. The binary variables $Z_{ic} \in \{0, 1\}$ indicates whether expert $i$ is active on cluster $c$.

Then we define some constraints to guarantee the legal mapping and efficiently search for the optimal allocation strategy.
\begin{align}
    & Z_{ic} \ge P_{ic},  \quad \forall (i,c) & \label{eq:const1}\\
    & \sum_{c} P_{ic} = 1,\quad  \forall i  & \label{eq:const4}\\
    & t_{\text{comp}} \ge \frac{\sum_{i=0}^{E-1}P_{ic}f_iB \cdot 2h \cdot IS}{\text{comp}},\quad  \forall c  & \label{eq:const2}\\
    & t_{\text{comm}} \ge \frac{4Bh}{\text{BW}} \sum_{g\in G} f_g\left(\sum_{i \in g} Z_{ic}\right) ,\quad  \forall c  & \label{eq:const3}\\
    & 0 < \sum_i^{E-1} P_{ic} f_i   \leq \left( \frac{1}{R_{CC}} + 1 \right) \frac{ e}{D},\quad  \forall c & \label{eq:const5} \\
    & R_{CC} = \frac{t_{\text{comp}}}{t_{\text{TP,comm}}} = \frac{\text{BW} \cdot IS \cdot e}{2D \cdot \text{comp}} & \label{eq:const6}
\end{align}

Constraints \ref{eq:const1} and \ref{eq:const4} handle expert placement and workload assignment across nodes. Constraints \ref{eq:const2} and \ref{eq:const3} ensure balanced computation and communication times, minimizing bottlenecks.

Constraints \ref{eq:const5} and \ref{eq:const6} restrict node workload to avoid imbalance, with an upper bound set by the theoretical compute + communication time of TP inference. These constraints prune suboptimal placements early, reducing search space complexity and improving solver convergence without sacrificing optimality.

Finally, we can represent and minimize the node-level inference overhead, defined as a combination of computation time and communication time:
\begin{align}
&\min t_{\text{node\_overhead}}\\
&t_{\text{node\_overhead}}=t_{\text{comp}}+2t_{\text{comm}}=t_{\text{comp}}+2\gamma \hat{t}_{\text{comm}}
\end{align}
Here, $\gamma$ is the scaling coefficient empirically derived from simulation (Sec. \ref{comm_model}), and the factor 2 accounts for the cost of all-to-all dispatch and all-to-all gather, which constitute a pair of symmetric communication operations.

This LP formulation enables a globally coordinated placement strategy that balances computation and communication, providing a foundation for the second stage of physical mapping.

\subsubsection{Link Balancing via Bayesian Optimization}\label{link balance}

In this stage, the logical clusters are mapped to physical nodes on the 2D mesh network. The objective here is to minimize link congestion and improve communication tail latency. We adopt Bayesian Optimization to search for low-congestion mapping strategies, as it is well-suited for problems with expensive evaluations and relatively smooth objective functions—e.g., swapping nearby clusters causes only minor changes in communication cost. This enables efficient exploration of the mapping space with minimal simulation overhead.

\begin{figure}[!tb]
\centerline{\includegraphics[width=\linewidth]{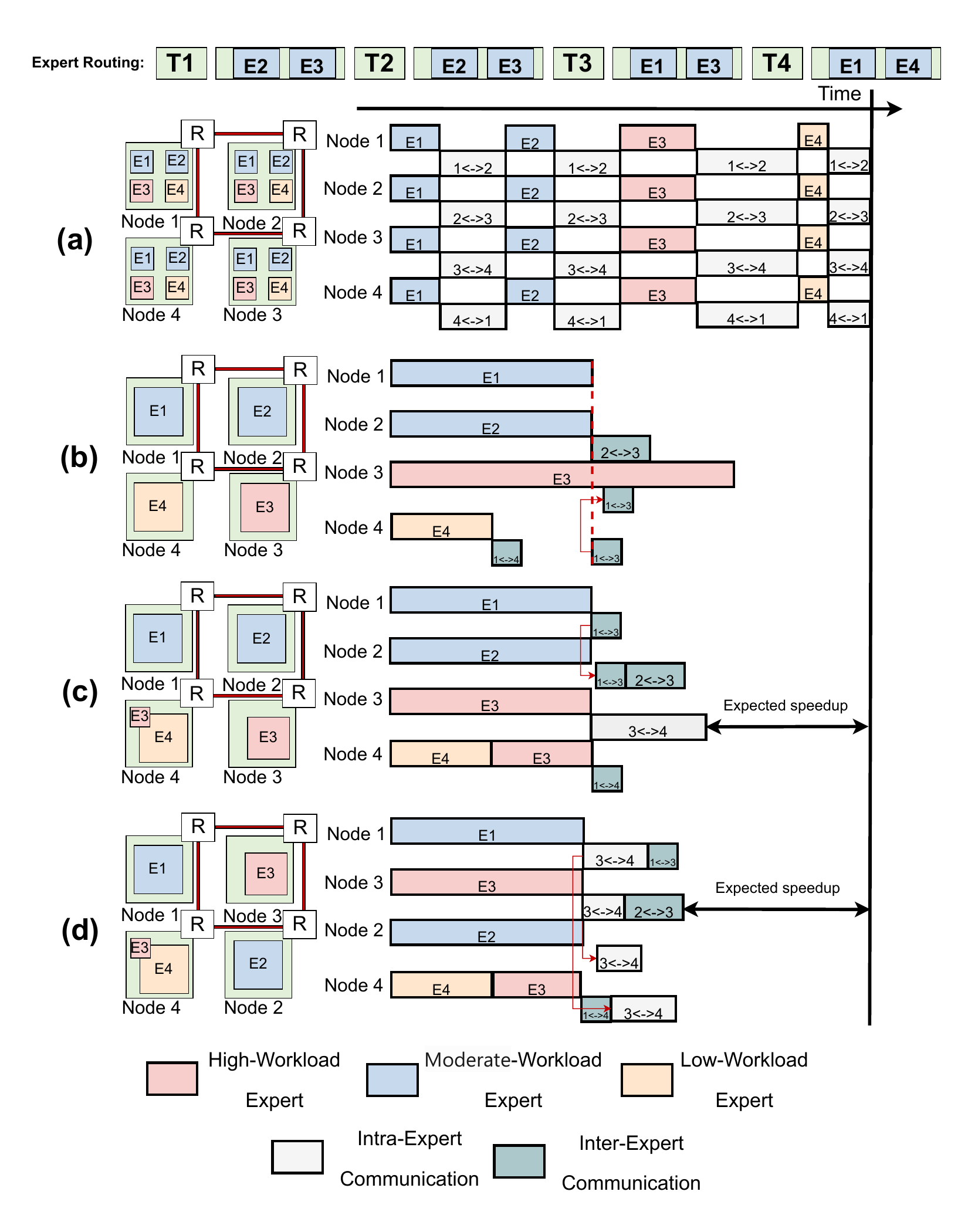}}
\caption{(a) TP: balanced computation but heavy communication; (b) EP: light communication but imbalanced computation, (c) Hybrid parallel with node balance: balanced computation but irregular communication; (d) Hybrid parallel with node-link balance: balanced computation and uncongested, regular communication.}
\label{node-link}
\end{figure}

Figure~\ref{node-link} illustrates example placements under four parallelism strategies, highlighting the trade-offs between computation balance and communication overhead in MoE inference. MoE communication can be categorized into Intra-Expert Communication, where a single expert is split across nodes and requires result aggregation (as in Tensor Parallelism), and Inter-Expert Communication, where multiple experts activated by the same token need to exchange and aggregate results. In (a), TP achieves balanced computation by splitting all experts across nodes, but incurs heavy Intra-Expert communication as each token triggers all-reduce operations. In (b), EP avoids Intra-Expert communication, with only sparse Inter-Expert communication—e.g., between E1 (Expert 1) and E3 via Node 4. However, tokens T1, T2, and T3 all activate E3, leading to overload on Node 3 and poor resource utilization. In (c), Node Balance optimization redistributes part of E3 to Node 4 to balance computation. Yet, this split introduces Intra-Expert communication between Nodes 3 and 4, causing link congestion along the 3→4 path while leaving other links underutilized. In (d), Node-Link Balance adjusts the physical mapping (e.g., swapping Node 2 and Node 3), allowing the synchronization between Nodes 3 and 4 to be routed via 3→1→4 and 3→2→4, which alleviates link congestion and achieves both balanced computation and regular, uncongested communication.

\subsection{Dynamic Placement Strategy}

We propose a runtime-adjustable deployment strategy for dynamic expert routing in MoE inference, consisting of three key components: congestion-aware expert prediction, cost-optimal broadcasting, and communication-efficient token routing.

\subsubsection{Priority Detection and Computation Prediction}\label{prediction}

We leverage temporal locality in expert activations to predict computation hotspots in the next layer. For each expert $i$ on node $c$, we define a priority score that estimates its future compute cost:
\begin{align}
&prio_{ic}=\frac{2P_{ic}\hat{f}_i\cdot IS}{comp}
\end{align}
Here, $\hat{f}_i$ is the predicted activation frequency of expert $i$. The expert with the highest priority on the most congested node is selected for pre-broadcast, repeating this process for a limited number of iterations based on the previous layer’s inference latency.

\subsubsection{Optimal Broadcast Chunk Size}\label{broadcast}

Broadcasting an expert involves splitting it into chunks of size $c$, with the following trade-offs:
\begin{itemize}
    \item Larger chunks reduce hops, lowering latency.
    \item Smaller chunks reduce per-hop traffic but increase transmission delays.
\end{itemize}

The traditional $\alpha$–$\beta$ communication model can clearly describe this kind of trade-off:
\begin{align}
&\text{latency}=\alpha (2\sqrt{D}+\frac{h\cdot IS}{c})\\
&\text{bandwidth}=\beta (h\cdot IS+2c\sqrt{D})\\
&t_{\text{pre\_b}}=\text{latency}+\text{bandwidth}
\end{align}
Here, $k$ is the number of pre-broadcast iterations allowed within the runtime window. The model yields a lower bound:
\begin{align}
    t_{\text{pre\_b}}\ge h\cdot IS\cdot\beta k+2\alpha \sqrt{D}+2\sqrt{2\sqrt{D}\beta k\alpha h\cdot IS}
\end{align}
This bound is tight when the chunk size $c$ is selected optimally as:
\begin{align}
    c=\sqrt{\frac{\alpha h\cdot IS}{2\beta k\sqrt{D}}}
\end{align}
This provides a solution for the most efficient pre-broadcast under a given runtime window constraint.
\subsubsection{Communication-Efficient Dispatch}\label{communication-eff}
After expert broadcasting, each token can be routed to any node holding a copy of its activated experts. To avoid incurring additional communication overhead, we restrict the routing candidates to nodes where the routed experts are already present. Among these candidates, the token is dispatched to the node with the lowest current compute load, minimizing workload imbalance without introducing extra data movement.

\begin{figure}[!tb]
\centerline{\includegraphics[width=\linewidth]{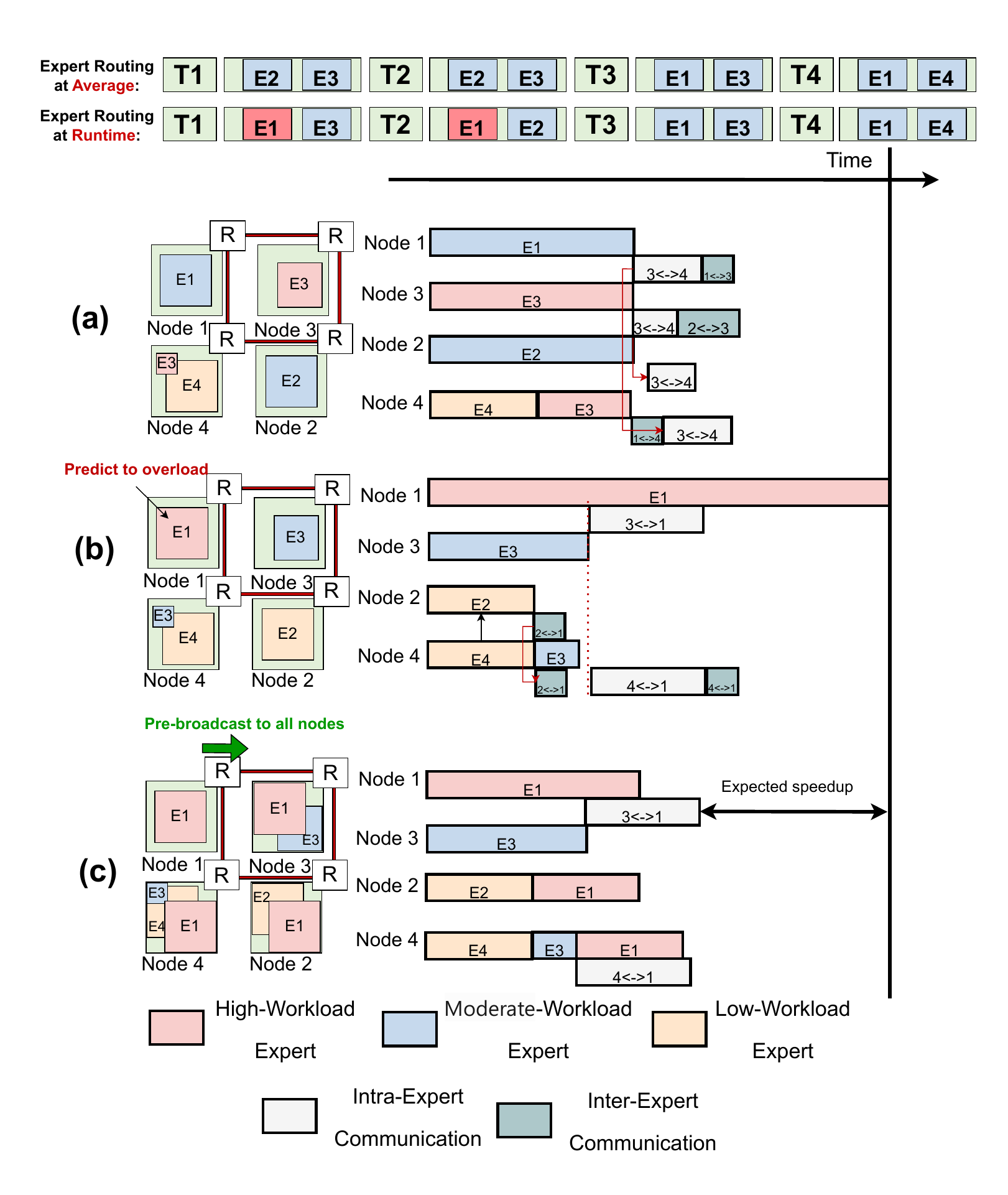}}
\caption{(a) Static deployment, (b) Computation load detection, (c) Pre-broadcast the expert with the highest load and dispatch tokens to appropriate nodes without inducing additional communication overhead.}
\label{fig:dynamic}
 \end{figure}

An example of dynamic expert scheduling is illustrated in Fig.~\ref{fig:dynamic}. Subfigure (a) shows the static deployment from Fig.~\ref{node-link}, which performs efficiently under averaged expert activation patterns. However, during real inference, as shown in (b), expert activation becomes highly skewed—Expert 1 (E1) turns into a bottleneck due to concentrated token routing. Our priority detection mechanism can anticipate such overload at runtime, identifying E1 as a high-demand expert in advance. As shown in (c), E1 is then \textbf{pre-broadcast} to all nodes before execution. Tokens such as T2 and T4 are routed to Node 2 and Node 4 respectively, both holding E1, thus avoiding additional Inter-Expert communication. This not only balances the computation load but also improves communication efficiency.
\section{Experimental Results}

\subsection{Experimental Setup}

\subsubsection{Models}
We evaluate the performance of our proposed approach using three MoE models: Mixtral-8x7B-Instruct \cite{jiang2024mixtral} (mixtral), DeepSeek-V2-Lite-Chat \cite{deepseekv2} (deepseek) and Qwen2-57B-A14B-Instruct \cite{yang2024qwen2technicalreport} (qwen). All the models are large-scale MoE architectures that benefit from expert parallelism and tensor parallelism, and they are deployed on 3D NMP architectures with different mesh sizes and hardware configurations. The key parameters for both models are summarized in Table \ref{model}.

\begin{table}[htbp]
\centering
\caption{Model Parameters for mixtral, deepseek and qwen}

\begin{center}
\resizebox{0.5\textwidth}{!}{ 
\begin{tabular}{c|c|c|c}

\hline\hline
\text{Parameter} & \text{mixtral} & \text{deepseek} & \text{qwen} \\
\hline
\text{Number of Experts} & 8 & 64 & 64 \\
\hline
\text{Experts per Token (Routing)} & 2 & 6 & 8 \\
\hline
\text{Number of Layers} & 32 & 27 & 28 \\
\hline
\text{Hidden Size} & 4096 & 2048 & 3584 \\
\hline
\text{Intermediate Size} & 14336 & 1408 & 2560 \\
\hline\hline
\end{tabular}}
\label{model}
\end{center}
\end{table}

\subsubsection{Baselines}
The baseline deployment strategies include Tensor Parallelism (TP), Expert Parallelism (EP), and a hybrid TP-EP approach with compute balance. In the hybrid strategy, the 2D mesh is divided into sub-regions—8 for deepseek and qwen, and 2 for mixtral. Each sub-region applies EP, with TP used internally to parallelize expert computation. Experts are appropriately assigned to sub-regions to balance computation load, with each expert placed in only one sub-region. This strategy is widely used to mitigate load imbalance in large-scale systems.

\subsubsection{Evaluation Metrics}
We evaluate the performance of our approach and the baselines using the following metrics:

\textbf{Normalized TBT (Time-Between-Tokens):} The latency between tokens during inference divided by that latency in Tensor Parallelism. 

\textbf{MoE Decomposed Latency:} The time taken to process a batch of tokens in MoE layers, including both computation and communication time.
\begin{itemize}
    \item Computation Latency: The time spent on performing computations within each node.
    \item Communication Latency: The time spent on transferring data between nodes.
\end{itemize}

\subsubsection{Dataset}
We use the MT Bench dataset \cite{10.5555/3666122.3668142} for evaluation, which is a widely adopted benchmark for LLMs, designed to measure the performance of LLMs on various tasks.

\subsubsection{Offline Optimization}
The proposed \textbf{Optimal Placement Strategy Searching} process typically takes \textbf{several hours} for the entire procedure, which is acceptable as it only needs to be performed once.

\subsection{End-to-End Performance}

In this section, we evaluate the end-to-end performance of the proposed Node-Link Balance strategy across different hardware configurations and 2D mesh sizes inferred in different batch sizes. The experiments are conducted using three hardware configurations, each with varying compute throughput and communication bandwidth:

\begin{itemize}
    \item 2.5 TFLOPS compute throughput, 75 GB/s bandwidth
    \item 5 TFLOPS compute throughput, 50 GB/s bandwidth
    \item 10 TFLOPS compute throughput, 25 GB/s bandwidth
\end{itemize}

Additionally, we compare the performance across three different 2D mesh sizes: (4,4), (4,8), and (8,8), with 5 TFLOPS compute throughput and 50 GB/s bandwidth for consistency.

\begin{figure*}[!tb]
    \centering
    \scriptsize
    \begin{tabular}{cccc}
    
        \textcolor{color1}{\rule{1em}{1em}} TP & 
        \textcolor{color2}{\rule{1em}{1em}} EP & 
        \textcolor{color3}{\rule{1em}{1em}} Compute Balance & 
        \textcolor{color4}{\rule{1em}{1em}} Node-Link Balance\\
    \end{tabular} \\[0.5em]
 
\centerline{\includegraphics[width=\linewidth]{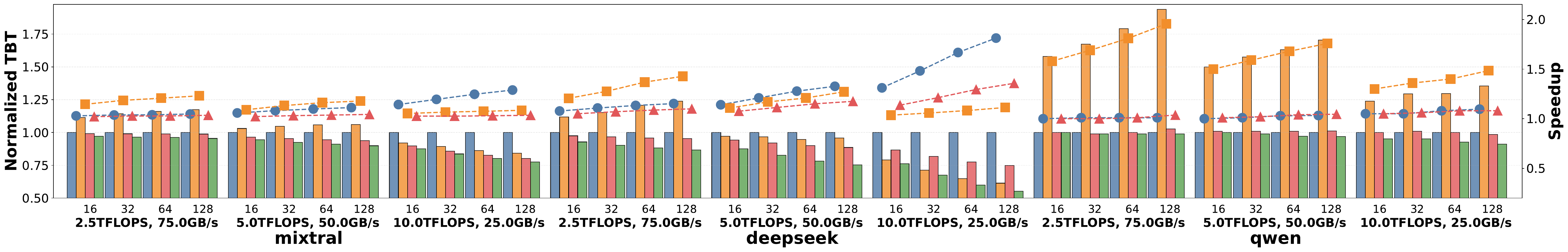}}
\caption{End-to-end Speedup for Different Hardware Configurations}
\label{config}
\end{figure*}

\begin{figure*}[!tb]
    \centering
    \scriptsize

\centerline{\includegraphics[width=\linewidth]{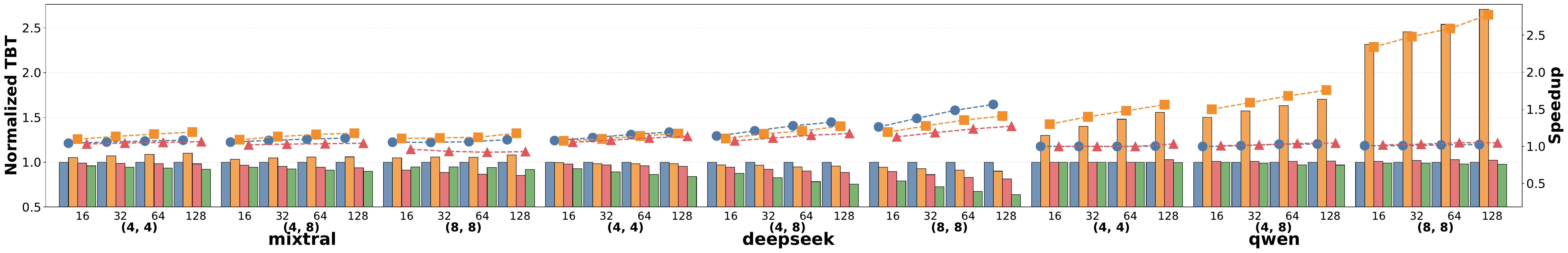}}
\caption{End-to-end Speedup for Different Mesh Shapes}
\label{mesh}
\end{figure*}

\textbf{Better TBT latency through different hardware configurations:} The results of these experiments, shown in Fig. \ref{config} and Fig. \ref{mesh}. Results in Fig. \ref{config} reveal how different methods respond to shifts in compute-to-communication ratios. When computation is limited and communication bandwidth is abundant (2.5 TFLOPS, 75 GB/s), EP suffers from severe workload imbalance, resulting in suboptimal TBT latency. In contrast, when computation is sufficient but communication becomes a bottleneck (10 TFLOPS, 25 GB/s), TP incurs heavy all-reduce communication costs, leading to degraded latency performance. Additionally, worth noting is that, for qwen, the expert routing exhibits high imbalance (Fig.~\ref{fig:motivation}(a)), causing significant overhead in EP.

The Hybrid TP-EP with Compute-Balanced baseline achieves better performance by distributing expert load more evenly, but ignores communication topology, leading to degraded performance under constrained bandwidth.

In contrast, our Node-Link Balance strategy jointly considers both computation and communication during expert placement. It minimizes per-node compute load, inter-node communication volume, and per-link congestion. As a result, it consistently outperforms all baselines across different system configurations.

On average, our method achieves a speedup ranging from \textbf{1.1$\times$ to 1.8$\times$} compared to TP, \textbf{1.1$\times$ to 1.5$\times$} compared to EP, and \textbf{1.0$\times$ to 1.4$\times$} compared to Hybrid TP-EP with Compute-Balanced.

\textbf{Better TBT latency through different mesh size:} We evaluate the impact of mesh topology on TBT latency under a fixed configuration (5 TFLOPS, 50 GB/s). As shown in Fig. \ref{mesh}, our Node-Link Balance strategy consistently achieves low latency across mesh sizes, demonstrating strong adaptability.

An exception occurs in mixtral with an (8,8) mesh, where the Hybrid TP-EP with Compute-Balanced baseline achieves slightly better latency. This is likely due to mixtral’s small number of experts, which must be spread across multiple nodes. In large mesh topologies, where communication regularity is more critical, the hybrid baseline benefits from its structured TP communication and moderate message volume.

Overall, our method remains effective across models and mesh sizes, particularly when the number of experts and the topology scale are well matched.

\subsection{Ablation Study}

We further conduct an ablation study focused on the contribution of the \textbf{Node Balance}, \textbf{Link Balance}, and \textbf{Dynamic scheduling optimization} for deepseek.

\subsubsection{Node Balancing}

\begin{figure}[!tb]
    \centering
    \scriptsize
    \begin{tabular}{cccc}
        \textcolor{color1}{\rule{1em}{1em}} TP & 
        \textcolor{color2}{\rule{1em}{1em}} EP & 
        \textcolor{color3}{\rule{1em}{1em}} Compute Balance & 
        \textcolor{color4}{\rule{1em}{1em}} Node-Link Balance\\
    \end{tabular} \\[0.5em]
\centerline{\includegraphics[width=\linewidth]{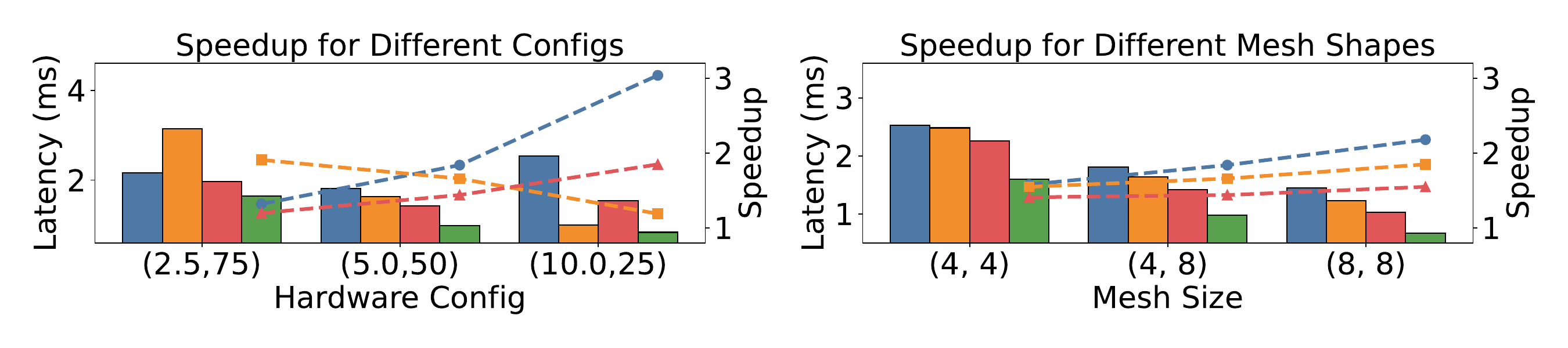}}
\caption{Node Balancing Speedup for DeepSeekMoE}
\label{node}
\end{figure}

We first evaluate the effect of the Node Balance stage in Fig. \ref{node}. It achieves \textbf{1.0$\times$ to 3.0$\times$} speedup over TP and EP, and \textbf{1.5$\times$} over Hybrid TP-EP across various configurations, by improving compute load balance (vs. EP) and reducing communication volume (vs. TP and hybrid).

\begin{figure}[!tb]
    \centering
    \scriptsize
    \begin{tabular}{cccc}
        \textcolor{color1}{\rule{1em}{1em}} EP & 
        \textcolor{color2}{\rule{1em}{1em}} Node-Link Balance & 

    \end{tabular} \\[0.5em]
\centerline{\includegraphics[width=\linewidth]{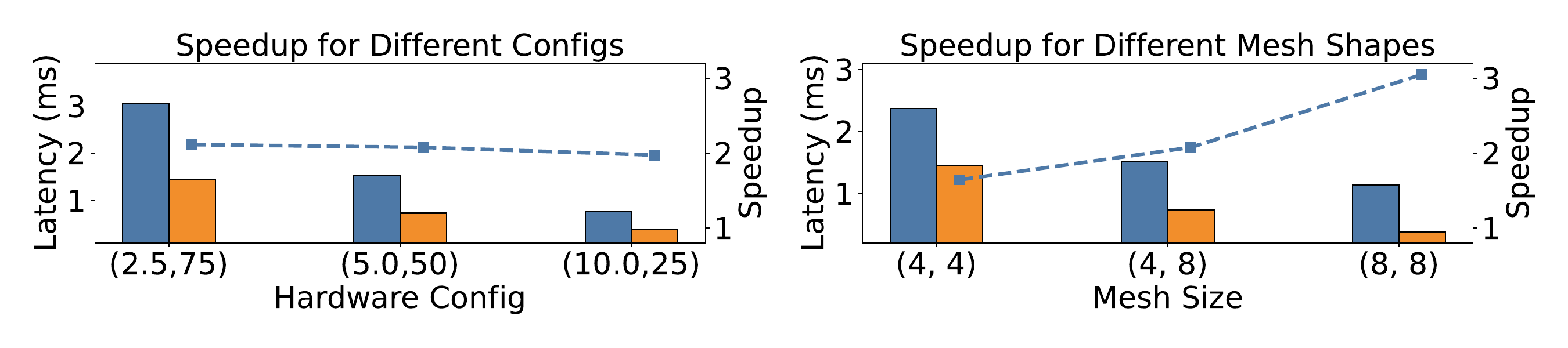}}
\caption{Node Balancing Speedup in Computation for DeepSeekMoE}
\label{node_comp}
\end{figure}

\textbf{Better computation latency:} Next, we specifically examine Node Balance's effect on compute imbalance in EP, by measuring its impact on computation latency within MoE layers As shown in Fig. \ref{node_comp}, on average, Node Balance reduces EP's compute tail latency by \textbf{2.0$\times$}, confirming its effectiveness in mitigating the routing skew commonly observed in MoE models.

\begin{figure}[!tb]
\centerline{\includegraphics[width=\linewidth]{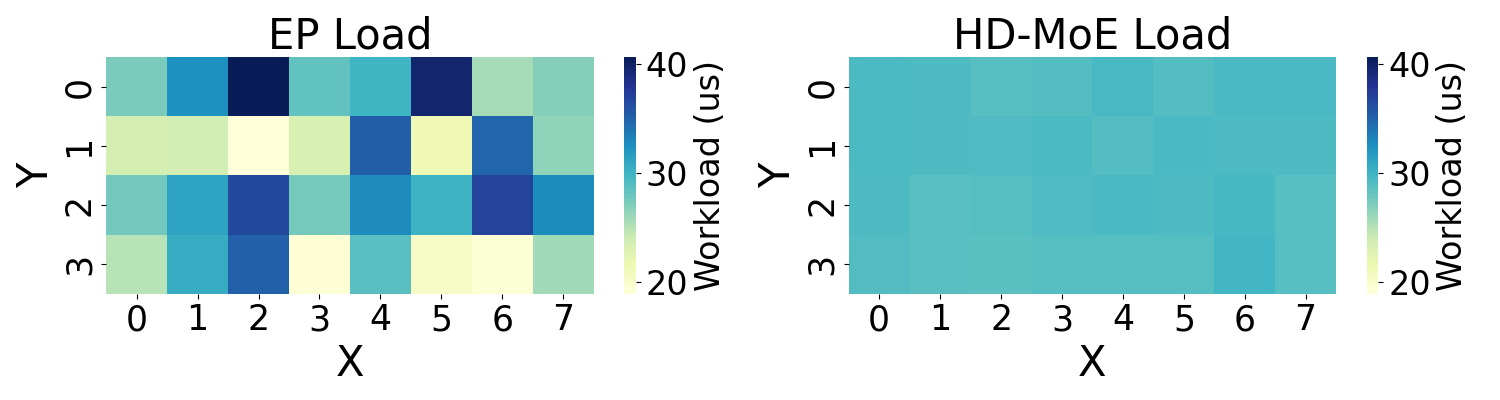}}
\caption{Visualization of Node-Level Resource Utilization With and Without Node Balance Optimization}
\label{load}
\end{figure}

\textbf{Better load balance:} Fig. \ref{load} shows per-node compute and communication load before and after applying Node Balance. The optimized placement achieves noticeably better load balance than standard EP, which often exhibits severe hotspots.

\subsubsection{Link Balancing}
\begin{figure}[!tb]
    \centering
    \scriptsize
    \begin{tabular}{cccc}
        \textcolor{color1}{\rule{1em}{1em}} TP & 
        \textcolor{color2}{\rule{1em}{1em}} EP & 
        \textcolor{color3}{\rule{1em}{1em}} Compute Balance & 
        \textcolor{color4}{\rule{1em}{1em}} Node-Link Balance\\
    \end{tabular} \\[0.5em]
\centerline{\includegraphics[width=\linewidth]{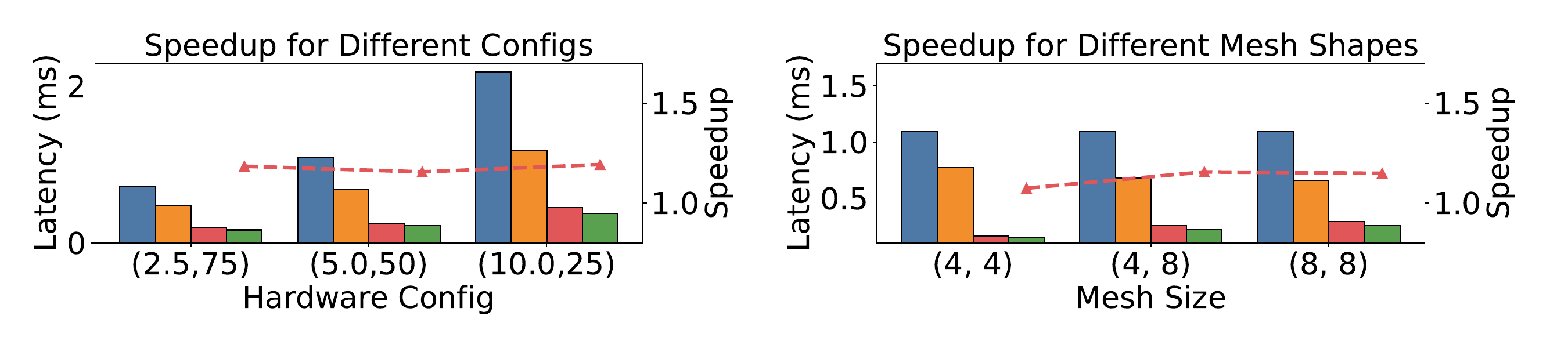}}
\caption{Link Balancing Speedup for DeepSeekMoE}
\label{link}
\end{figure}

We further evaluate the contribution of the Link Balance stage by isolating its impact on communication latency in Fig. \ref{link}. Specifically, we compare against three baselines: TP, Hybrid TP-EP with Compute-Balanced, and the Node Balance without physical mapping optimization.

\textbf{Better communication latency:} Thanks to the Bayesian Optimization–based mapping strategy, Link Balance produces more communication-friendly mappings by assigning logical clusters to physical nodes in a topology-aware manner. This significantly reduces link congestion and results in lower communication latency than TP and hybrid baselines, which rely on regular but heavy communication.

Compared to the Node Balance–only deployment, Link Balance can also achieve an average \textbf{1.2$\times$} reduction in communication latency, highlighting the importance of mapping logical clusters to physical nodes with awareness of network structure.

\begin{figure}[!tb]
\centerline{\includegraphics[width=\linewidth]{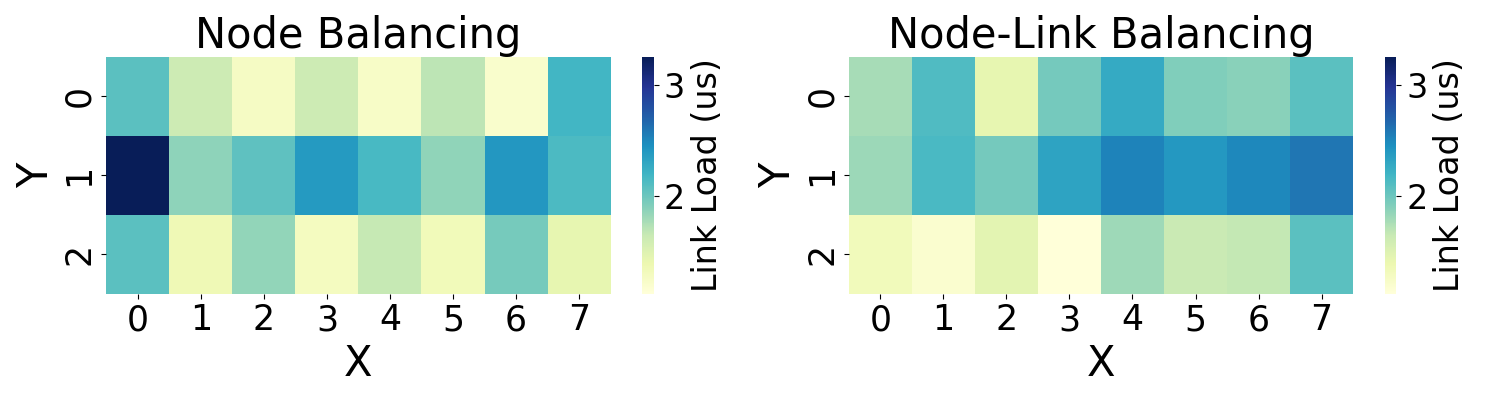}}
\caption{Visualization of Link-Level Resource Utilization With and Without Link Balance Optimization}
\label{link_eg}
\end{figure}

\textbf{Less link congestion:} Fig. \ref{link_eg} visualizes NoC link utilization using heatmaps. Compared to the Node Balance stage, the optimized placement after Link Balance leads to visibly more balanced link-level traffic, with less link congestion and better distribution across the mesh.

\begin{figure}[!tb]
\centering
\scriptsize
    \begin{tabular}{cccc}
        \textcolor{color1}{\rule{1em}{1em}} Static Deployment & 
        \textcolor{color2}{\rule{1em}{1em}} Dynamic Deployment & 

    \end{tabular} \\[0.5em]
\begin{subfigure}[t]{0.24\textwidth}
    \includegraphics[width=\linewidth]{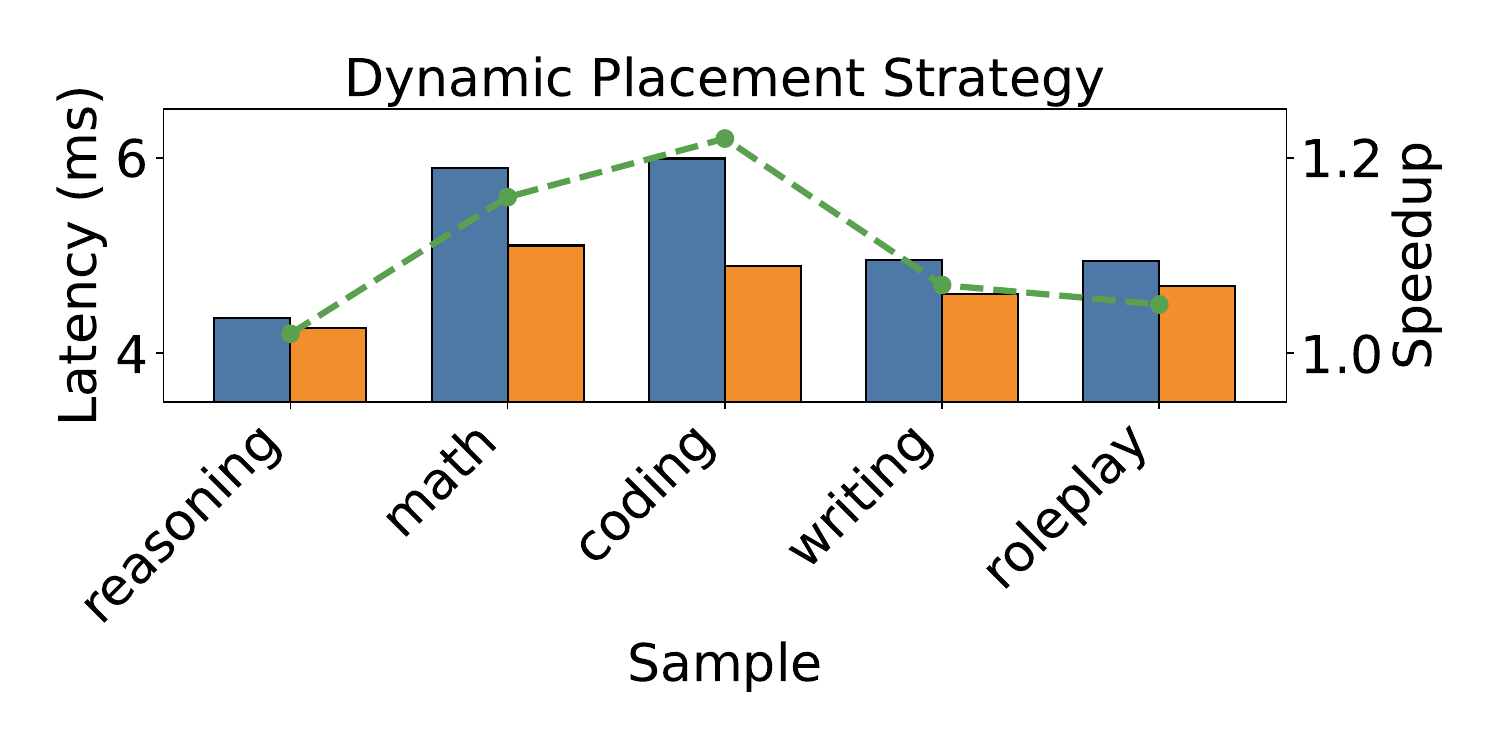}
    \caption{}
    \label{dynamic2}
\end{subfigure}
\hfill
\begin{subfigure}[t]{0.24\textwidth}
    \includegraphics[width=\linewidth]{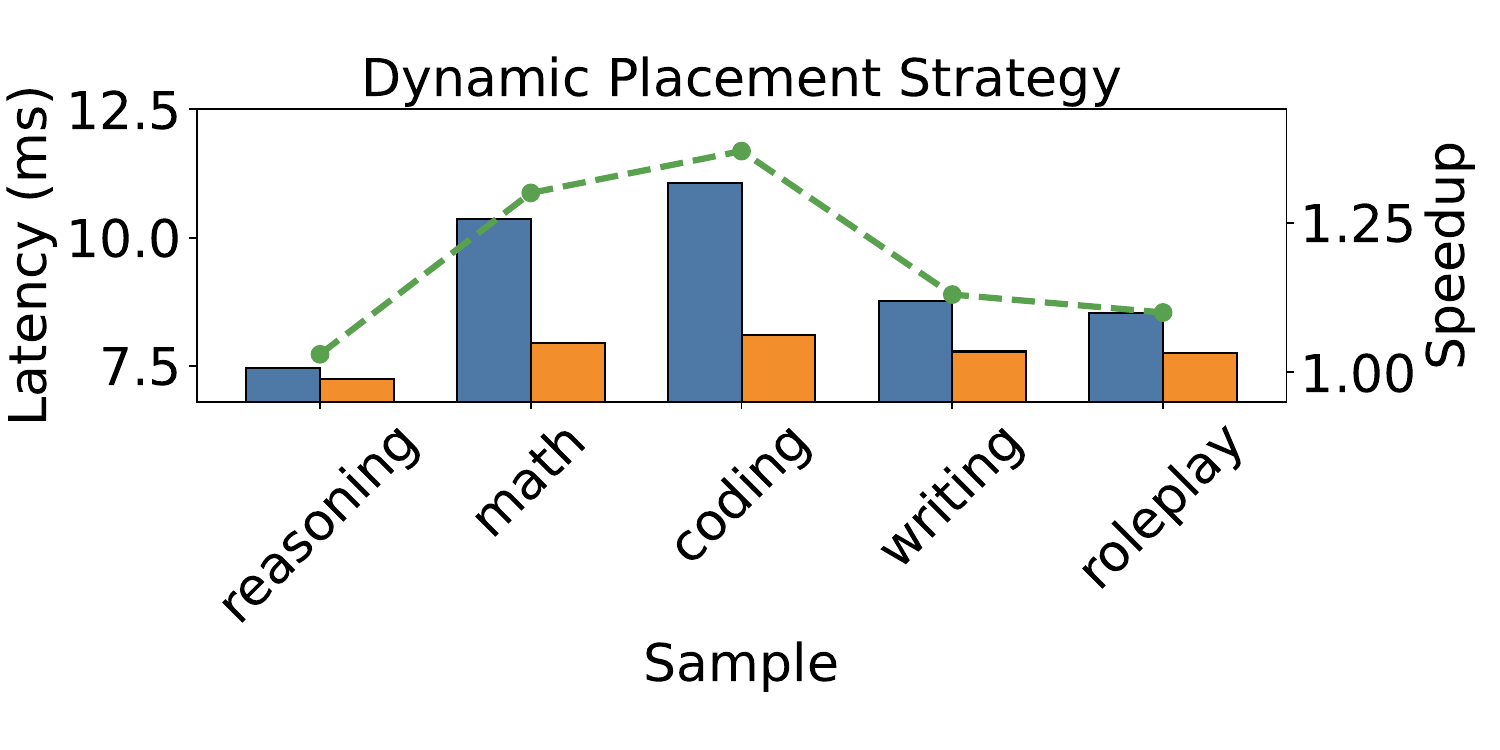}
    \caption{}
    \label{dynamic5}
\end{subfigure}
\caption{Latency and Speedup Comparison Between Static and Dynamic Placement Strategies Under Varying Inference Scenarios. (a) Pre-broadcast 2 experts, (b) Pre-broadcast 5 experts.}
\label{dynamic}
\end{figure}

\subsubsection{Dynamic Placement Strategy}

To assess the impact of the Dynamic Placement Strategy, we compare the performance of static and dynamic expert placement strategies. In these experiments, we sample multiple expert routing traces in various types of questions from the MT Bench dataset, focusing on tasks with varying expert activation patterns. We compare the latency and speedup between static (generating from reasoning questions) and dynamic strategies under two different hardware configurations and broadcasting settings:

Hardware Configuration: (5 TFLOPS, 50 GB/s bandwidth) with 512 batch size, which has enough time to pre-broadcast 2 experts per layer, the results are shown in Fig. \ref{dynamic}(a).

Hardware Configuration: (2.5 TFLOPS, 75 GB/s bandwidth) with 512 batch size, which has enough time to pre-broadcast 5 experts per layer, the results are shown in Fig. \ref{dynamic}(b).

\textbf{Better performance in various scenarios:} The results are shown in Fig. \ref{dynamic}, which indicates that the Dynamic Placement Strategy provides significant speedups and maintains relatively stable inference latency across a variety of real-time inference scenarios. Notably, for tasks such as math and coding problems, which have huge differences from reasoning, the dynamic approach significantly reduces MoE layer latency compared to static deployments. Specifically, when broadcasting 2 experts per layer, the average speedup achieved by the dynamic strategy is 1.15$\times$, and when broadcasting 5 experts per layer, the average speedup increases to 1.25$\times$.

These results highlight the effectiveness of dynamic expert scheduling in reducing latency by adapting to inference-time workload and improving both computation and communication efficiency.

\section{Conclusion}

This paper presents \textbf{HD-MoE}, an offline \textbf{Automatic Hybrid Parallelism} strategy, combined with online \textbf{Dynamic Scheduling}, for efficiently deploying MoE models on 3D NMP architectures. By integrating Node Balance, Link Balance, and Dynamic Placement, our approach effectively reduces computation and communication latency in MoE layers, improving both load balancing and resource utilization. Experimental results show that our method outperforms baseline strategies, achieving a speedup ranging from \textbf{1.1$\times$ to 1.8$\times$} over TP and \textbf{1.1$\times$ to 1.5$\times$} over EP. These findings demonstrate the value of optimizing expert placement and dynamic scheduling for MoE deployment on NMP architectures.

\newpage

\bibliographystyle{IEEEtran}
\bibliography{reference}

\end{document}